\documentclass[journal]{IEEEtran}
\usepackage{multirow}

\ifCLASSINFOpdf
  \usepackage[pdftex]{graphicx}
  \graphicspath{{./figures/}}
  \DeclareGraphicsExtensions{.pdf,.jpeg,.png}
\else
  \usepackage[dvips]{graphicx}
  \graphicspath{{./figures/}}
  \DeclareGraphicsExtensions{.eps}
\fi
\begin{document}
%
\title{The MicroTCA fast control board \\
for generic control and data acquisition applications \\
in HEP experiments}
%
%
\author{Jie~Zhang, Hangxu~Li, Jingzi~Gu, Xiaoshan~Jiang
\thanks{The authors are with State Key Laboratory of Particle Detection and Electronics, Institute of High Energy Physics, CAS, Beijing 100049 China e-mail: zhj@ihep.ac.cn}
\thanks{Manuscript received June 24, 2018.}}
%
%

\markboth{IEEE TRANSACTION ON NUCLEAR SCIENCE,~Vol.~x, No.~x, June~2018}%
{Shell \MakeLowercase{\textit{et al.}}: IEEE TRANSACTION ON NUCLEAR SCIENCE}
%



\maketitle

\begin{abstract}
    To provide the generic clock, trigger and control function and fulfill the data transmission performance requirement in high energy physics (HEP) experiments, a MicroTCA Fast Control board (uFC) was developed based on the Advanced Mezzanine Card (AMC) specification. Built around the Xilinx Kintex-7 FPGA, the uFC provides users with a platform with data memory, reference clock, trigger and SFP+ (Small Form-factor Pluggable) connections that are required in general experiment. In addition, it has access to two on-board FPGA Mezzanine Card (FMC) sockets with a large array of configurable I/O and high-speed links up to 10 Gbps. This paper presents test results from the first set of pre-production prototypes and reports on the application in High Energy Photon Source in China.
\end{abstract}

\begin{IEEEkeywords}
high energy physics experiments, modular electronics, MicroTCA, AMC, FMC.
\end{IEEEkeywords}

%
\IEEEpeerreviewmaketitle

\section{Introduction}

\begin{figure*}[!hbtp]
\centering
\includegraphics[width=2.5in]{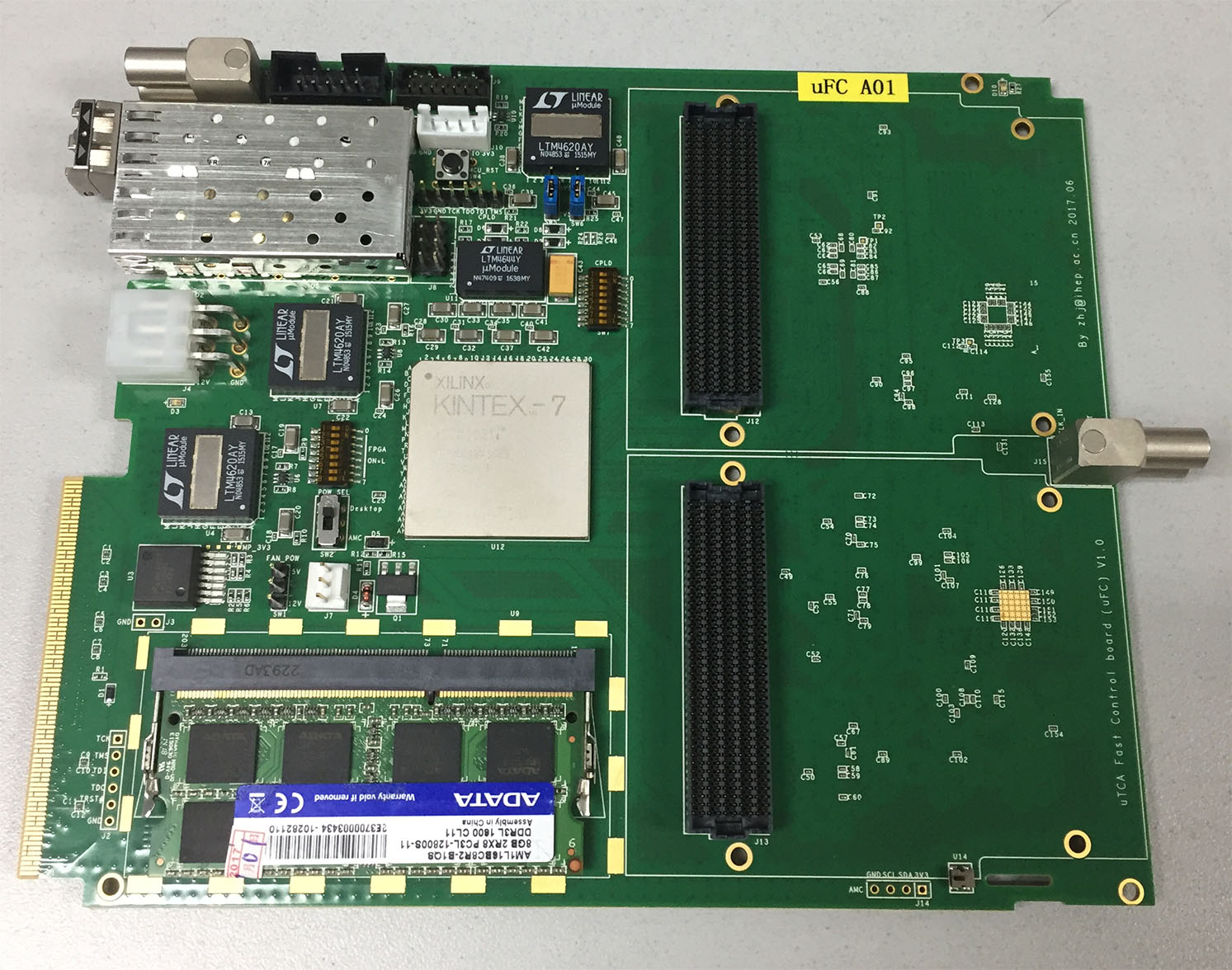}
\hfil
\includegraphics[width=2.5in]{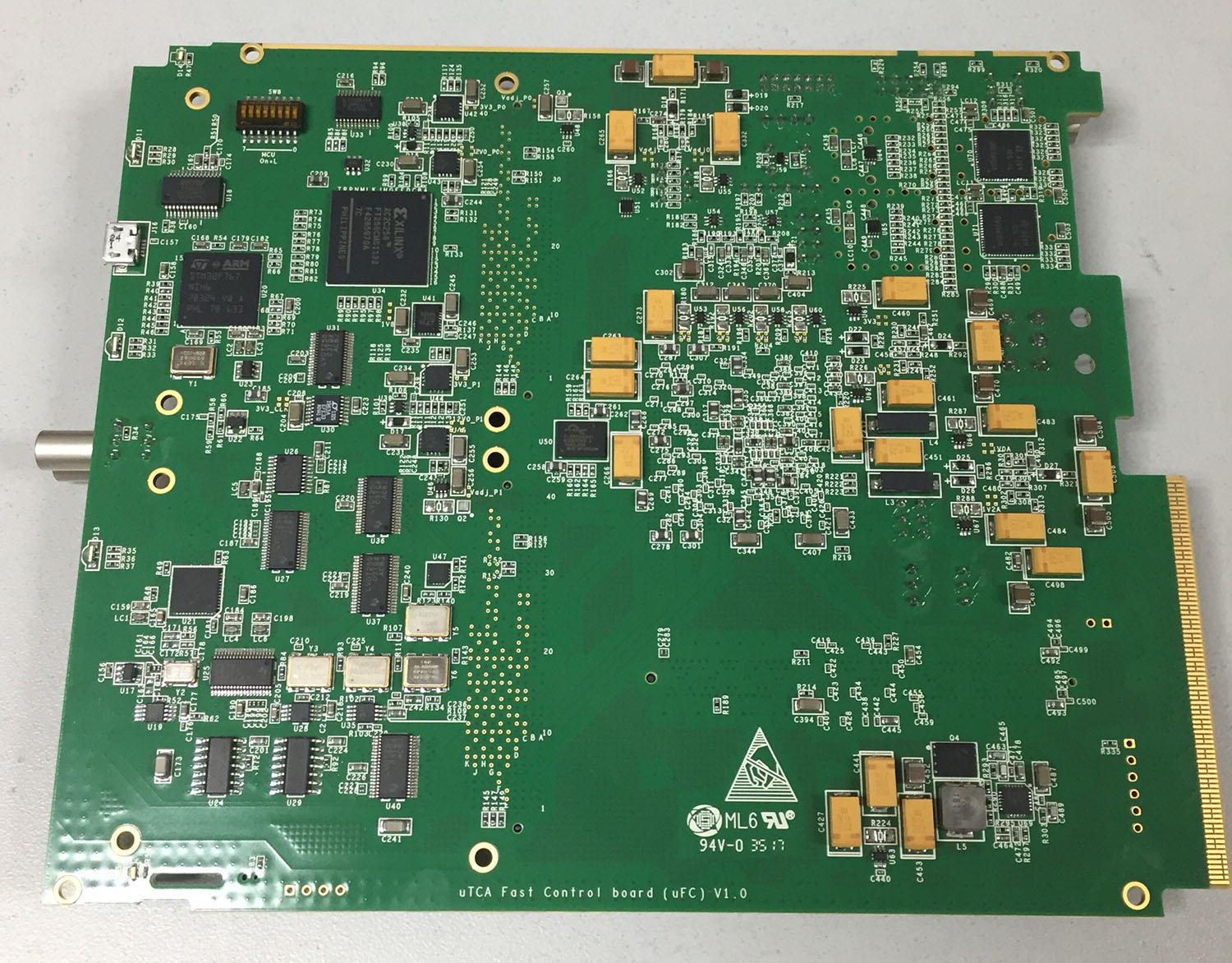}
\caption{Front (left) and back (right) pictures of the uFC with the Xilinx Kintex-7 FPGA and two HPC compatible FMC sockets for generic input and output. In the left picture, the AMC edge connector is located on the left side of the board while the front panel is on the right.}
\label{fig_ufc}
\end{figure*}

\IEEEPARstart{T}{he} Micro Telecommunications Computing Architecture (MicroTCA) standard\cite{MicroTCA} is a modern platform that is gaining popularity in the area of HEP experiments\cite{Physics}\cite{CMS}. The main advantage of this architecture is high-level reliability, availability and maintainability. The concept for developing the uFC is an FPGA-based development platform designed to serve general purpose control and data acquisition system residing either inside a MicroTCA crate or stand-alone on a bench with high-speed optical/Ethernet links. Figure \ref{fig_ufc} shows a picture of the pre-production uFC, highlighting two SFP+ connectors, hard gold AMC edge connector, two high-pin count (HPC) compatible FMC\cite{FMC} sockets and 8GB DDR3 SODIMM (small outline dual in-line memory module). By basing the uFC on existing hardware, extracting commonality between projects, the development time for experiment can be short by reusing design components, including hardware, firmware and software.

This paper has the following structure: Section II and III describe the hardware and firmware design of the uFC. In Section IV, we compare the hardware scheme with two other double-width FPGA-based AMC with dual-FMC. The performance evaluation and the first results from the application in High Energy Photon Source (HEPS) are demonstrated in the Section V.


\section{Hardware}
\begin{figure}[!htbp]
\centering
\includegraphics[width=2in]{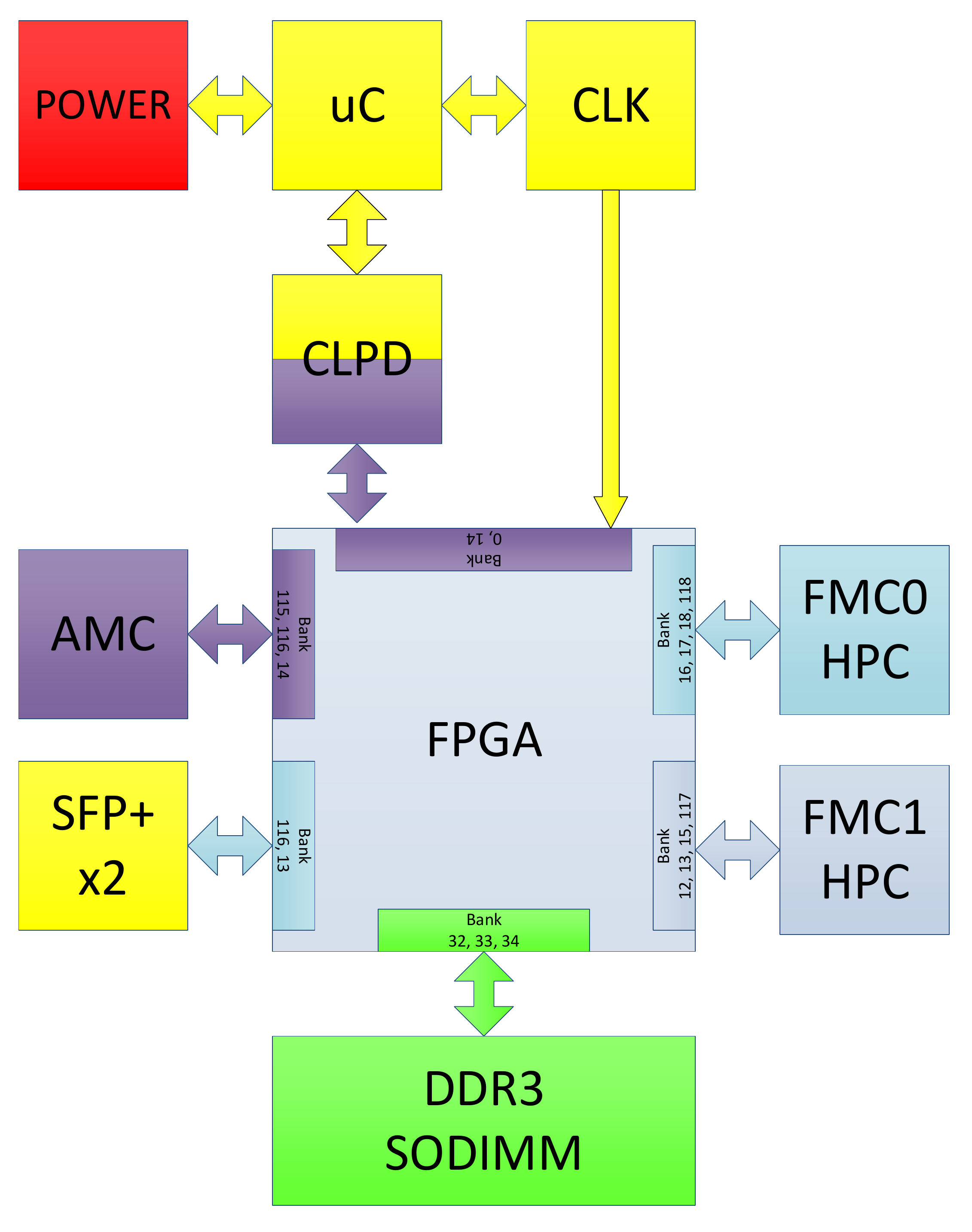}
\caption{The block diagram of the uFC}
\label{fig_ufc_block}
\end{figure}

Designed as a middle-size, double-width AMC, the uFC is suitable for MicroTCA-based scalable system, as well as for bench-top prototyping. Figure \ref{fig_ufc_block} shows the block diagram of the uFC.

\subsection{FPGA}
Based on the Xilinx Kintex-7 XC7K325T-2FFG900I FPGA, the application allows designing a cost-effective and still powerful controller supporting sixteen links with rate up to 10 Gbps.

\subsection{Memory}
\begin{itemize}
\item Up to 8G-Byte DDR3 SODIMM with 64-bit wide data bus, capable of data transfer rate up to 64 Gbps at 500 MHz.
\item 32M-Byte SPI Flash Memory for storing the FPGA firmware bitstreams.
\item 2K-bit IIC EEPROM with EUI-48\textsuperscript{TM} Node Identity, providing a unique node Ethernet MAC address for mass-production process.
\end{itemize}

\subsection{Communication \& Networking}
Considering board to board transmission, the FPGA should communicate with other devices using high-speed point to point connections provided by the serial transceivers. The applied Kintex-7 FPGA allows a maximum throughput up to 10.3125 Gbps (speed grade -2). But unfortunately, it provides a hardware IP block for the PCI-Express 2.0 with 5 GT/s. This may upgrade to the PCI-Express 3.0 by a software IP solution. The uFC supports 1/10 Gb Ethernet connection on port 0 and 1, and PCI-Express x4 on port 4-7 for MicroTCA backplane. Taking advantage of the flexibility of the FPGA, the uFC gives users the possibility of implementing various other high-speed protocols instead of PCI-Express. Besides, the uFC carries two SFP+ connectors in order to interface to other board through optical link or to a PC through 1000BASE-T SFP transceiver module\cite{SFP_GE_T} and a category 5 cable while in bench-top operation.

\subsection{Expansion Connectors}
The uFC input/output capability can be further enhanced with two HPC-FMC sockets. Each has 4 GTX transceivers, 116 single-ended or 58 differential user defined signals (34 LA \& 24 HA). The V$_{adj}$ supports 1.8V, 2.5V, or 3.3V. These sockets allow the use of a variety of mezzanines for various applications, for instance by mounting ADC/DAC module for front-end, or connecting clock, trigger and control module to back-end. Moreover, the front and rear panels provide four LEMO input/output connectors for external clock and trigger function.

\subsection{Module Management Controller (MMC)}
ARM Cortex-M7 microcontroller is used as the MMC for the board initialization when in a MicroTCA crate as well as for other management tasks, e.g. temperature, voltage, current monitoring and FPGA control (payload reset, firmware reload, and check status of booting). The original firmware of the MMC was written by CERN MMC project\cite{MMC}. It is compliant with the Intelligent Platform Management Interface v1.5 standard as well as the PICMG extension for xTCA. We migrated this project from Atmel AVR to ARM Cortex-M7. The microcontroller monitors and controls all DC/DC converters via PMBus with the help of two UCD90120A chips from Texas Instruments. If the current of the FMC module exceeds the limitation, the UCD90120A will turn off power for FMC automatically and report to the MMC. The temperatures are measured under the DC/DC converters and also directly on the FPGA die. The MMC supports the firmware upgrade of the microcontroller and the FPGA using the HPM.1 protocol. The MMC firmware can be debugged using a virtual serial port, available through USB connection.

\subsection{Clocking}
The uFC offers input clock sources (Figure \ref{fig_clock}):

\begin{figure*}[!htbp]
\centering
\includegraphics[width=6in]{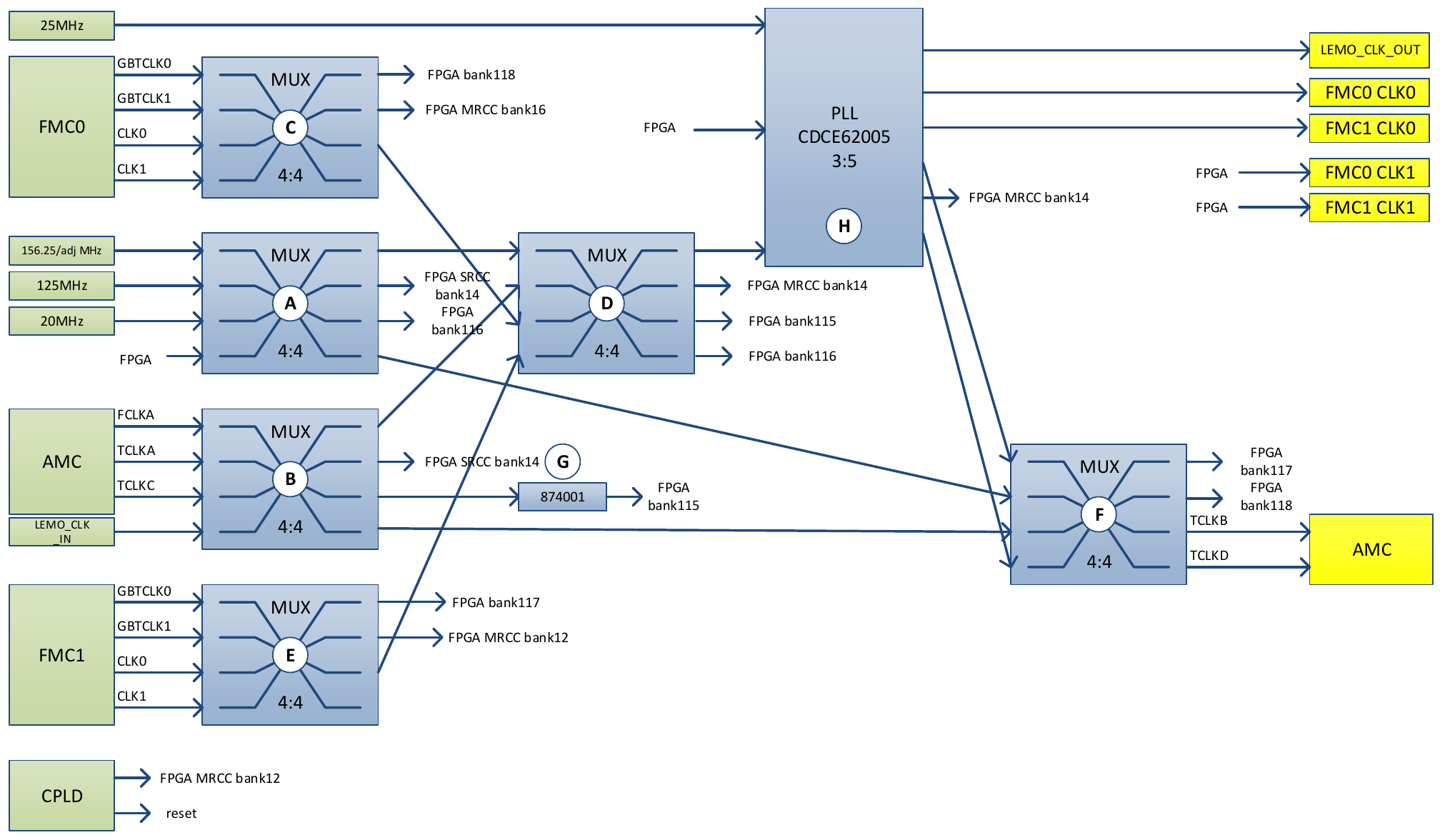}
\caption{Clock generation and distribution in uFC}
\label{fig_clock}
\end{figure*}

\begin{itemize}
\item Fixed Oscillator with single-ended 25 MHz through CPLD (Complex Programmable Logic Device).
\item Fixed Oscillator with differential 125 MHz for 1G Ethernet.
\item Fixed Oscillator with differential 200 MHz for FPGA IO Delay.
\item Programmable Oscillator with 156.250 MHz as the default output. Default frequency targeted for 10G Ethernet applications but oscillator is programmable for other end uses.
\item LEMO external clock input and output in front and rear panels.
\item AMC FCLKA, TCLKA and TCLKC input. TCLKB and TCLKD output.
\item FMC CLKx\_C2M, CLKx\_M2C and GBTCLKx\_M2C clocks.
\item VCXO and TCVCXO support the White Rabbit clocking which provides sub-nanosecond synchronization.
\item Jitter attenuated clock by programmable clock multiplier.
\end{itemize}
Based on cross-point switches and programmable clock multipliers, the clock distribution circuit offers a large selection of input clock sources (e.g. the LEMO connectors in the front/rear panel, the AMC clocks, the FMC clocks, or onboard oscillators). This makes the uFC give users the possibility of implementing various high speed serial data protocols for custom applications.

\subsection{External Power}
In bench-top prototyping, a 12V adapter is used as input power, and a switch helps to bypass the AMC initialization in MMC and 3.3V Management Power.

\subsection{Configuration}
\begin{figure}[!htbp]
\centering
\includegraphics[width=2.5in]{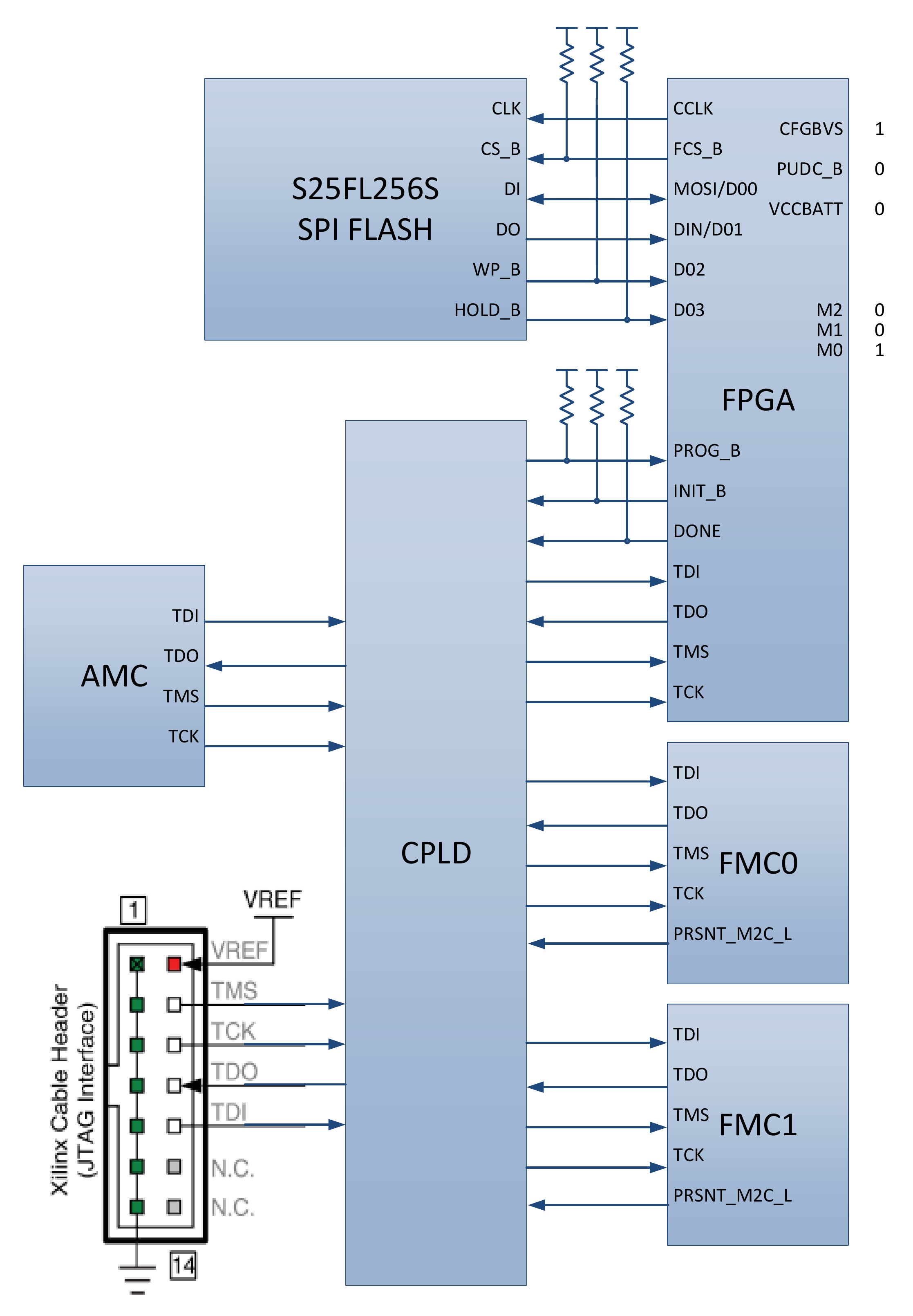}
\caption{JTAG programming connections in the uFC}
\label{fig_jtag}
\end{figure}

Xilinx allows configuring and debugging the FPGA through JTAG with download cables such as the Platform Cable USB II or Digilent USB cable. Users can access to the FPGA through the MicroTCA crate or JTAG header. The CPLD in the uFC acts as a bridge selecting the JTAG master source between the JTAG header and AMC JTAG lines (shown in Figure \ref{fig_jtag}). When an FMC card is attached to the uFC, CPLD automatically adds the attached device to the JTAG chain as determined by the FMC\_HPC\_PRSNT\_M2C\_B signal. The attached FMC card must implement a TDI to TDO connection via a device or bypass jumper for the JTAG chain to be completed.

\section{Firmware}
In order to simplify the application-specific development process for users, the uFC delivered a system core implemented all basic functions such as the Gigabit TCP/UDP Ethernet link based on SiTCP\cite{SITCP}, the MMC communication, the control of the clock circuitry and the interface with the DDR3 memory and I2C/SPI device.

\section{Compare with other FPGA-based AMC with dual-FMC}
With the popularization and development of MicroTCA, various FPGA-based AMCs with dual-FMC have been used in HEP experiments (e.g. LHC, E-XFEL, J-PARC). DAMC-FMC25 is one which developed by DESY\cite{DAMC-FMC25-DESY}, transformed into commercial product by CAENels\cite{DAMC-FMC25-CAENels}. Its fast links are also dedicated to the MTCA.4 standard use of the board with 42 differential pairs and 2 GTX \@ 6.5 Gbps to the RTM Zone-3 connector. FC7\cite{FC7}, Built upon the success of existing hardware developments - the Gigabit Link Interface Board (GLIB)\cite{GLIB}, is a new generation AMC for generic DAQ and control applications in CMS. Table \ref{tab_comparison} summarizes the differences and similarities between DAMC-FMC25, FC7 and uFC. We chose smaller FPGA considering the tradeoffs between cost and performance, replaced the RTM with SFP+ and LEMO connectors for bench-top prototyping.

\begin{table*}[!htbp]
\centering
\caption{Comparison Tables for DAMC-FMC25, FC7 and uFC}
\label{tab_comparison}
\begin{tabular}{|c|c|c|c|c|}
\hline
\multicolumn{2}{|c|}{} & DAMC-FMC25 & FC7  & uFC \\ \hline
\multicolumn{2}{|c|}{FPGA}  & \begin{tabular}[c]{@{}c@{}}XC5VFX70T/\\ XC5VFX100T\\ 2FFG1136\end{tabular} & \begin{tabular}[c]{@{}c@{}}XC7K420T\\ \\ FFG1156\end{tabular}        & \begin{tabular}[c]{@{}c@{}}XC7K325T\\ \\ FFG900\end{tabular}              \\ \hline
\multicolumn{2}{|c|}{Memory}  & 256MB DDR2 & 0.5GB DDR3 & Up to 8GB DDR3 SODIMM \\ \hline
\multirow{2}{*}{\begin{tabular}[c]{@{}c@{}}FMC\\ x2\end{tabular}} & IO   & 68  & 68   & 116 \\ \cline{2-5}
                                                                  & MGT  & 2/4 & 12/8 & 4/4 \\ \hline
\multirow{3}{*}{Communication}                                    & SFP+ & -   & -    & 2   \\ \cline{2-5}
                                                                  & \begin{tabular}[c]{@{}c@{}}AMC high-speed\\  connectivity\end{tabular} & \begin{tabular}[c]{@{}c@{}}Port 0, 1, 4$\sim$7, 12$\sim$15\\ Class D.1. for RTM\end{tabular} & \begin{tabular}[c]{@{}c@{}}Port 0$\sim$11\\ Without RTM\end{tabular} & \begin{tabular}[c]{@{}c@{}}Port 0, 1, 4$\sim$7\\ Without RTM\end{tabular} \\ \cline{2-5}
                                                                  & LEMO/SMA & 1  & 2  & 4  \\ \hline
\multicolumn{2}{|c|}{White Rabbit (WR)}                           & -   & -  & Yes    \\ \hline
\end{tabular}
\end{table*}

\section{Example use in HEPS}
\begin{figure}[!htbp]
\centering
\includegraphics[width=2.5in]{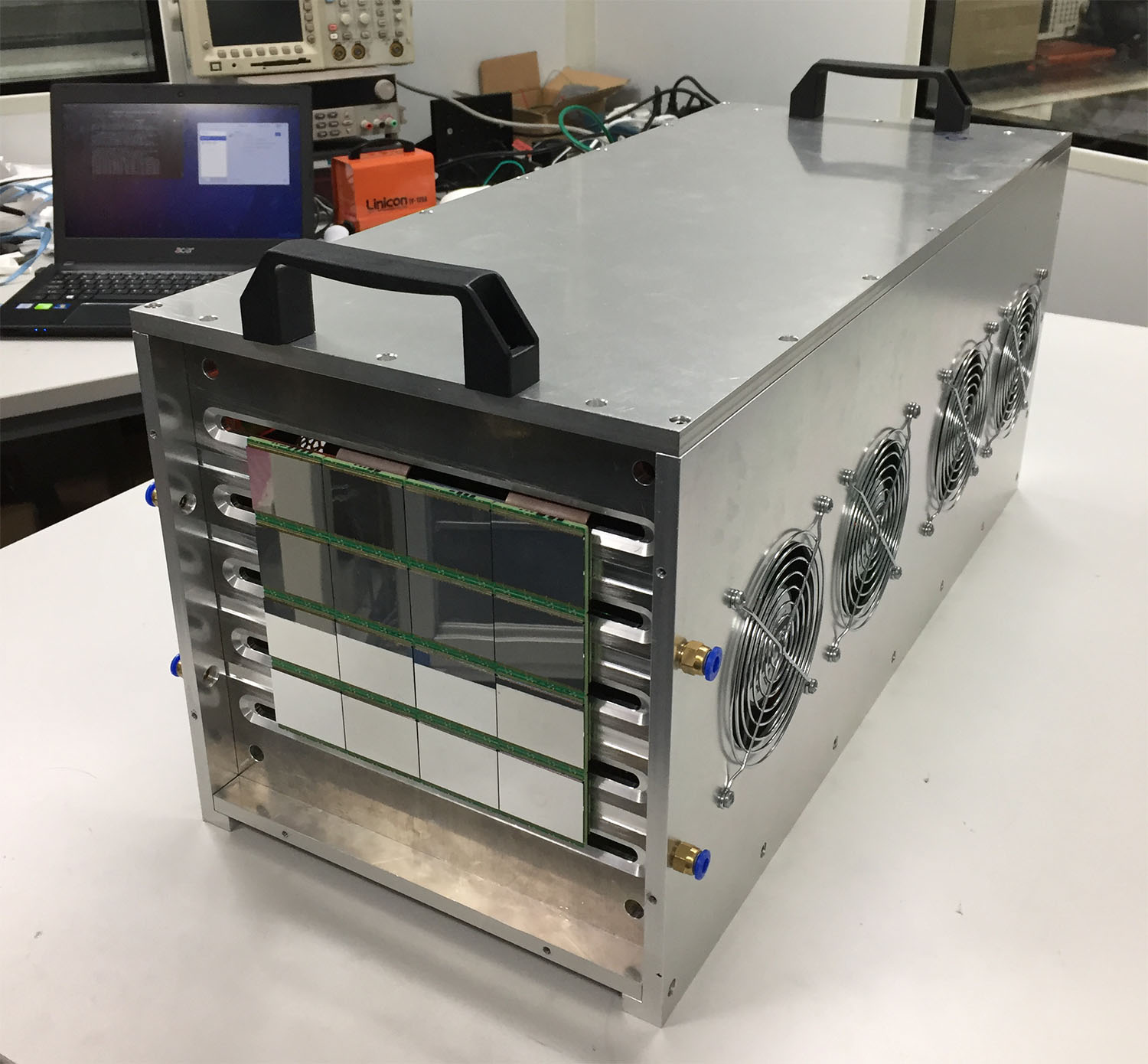}
\caption{Photo of the second generation HEPS-BPIX detector.}
\label{fig_bpix}
\end{figure}

The flexibility of the uFC with its variety of interfaces enables it to be used in various configurations. We adopted it as part of the second generation hybrid pixel detector (HEPS-BPIX)\cite{HEPS-BPIX} for the HEPS in China. This prototype system was assembled with sixteen front-end modules including 1M pixels in total, covering an area of 16.32 cm x 18.3 cm (Figure \ref{fig_bpix}). The data acquisition was provided by a single server through four 10G Ethernet, achieving a data rate of 2.3 GB/s in all at 16-bit, 1.2 kHz frame rate.

\begin{figure}[!htbp]
\centering
\includegraphics[width=2.5in]{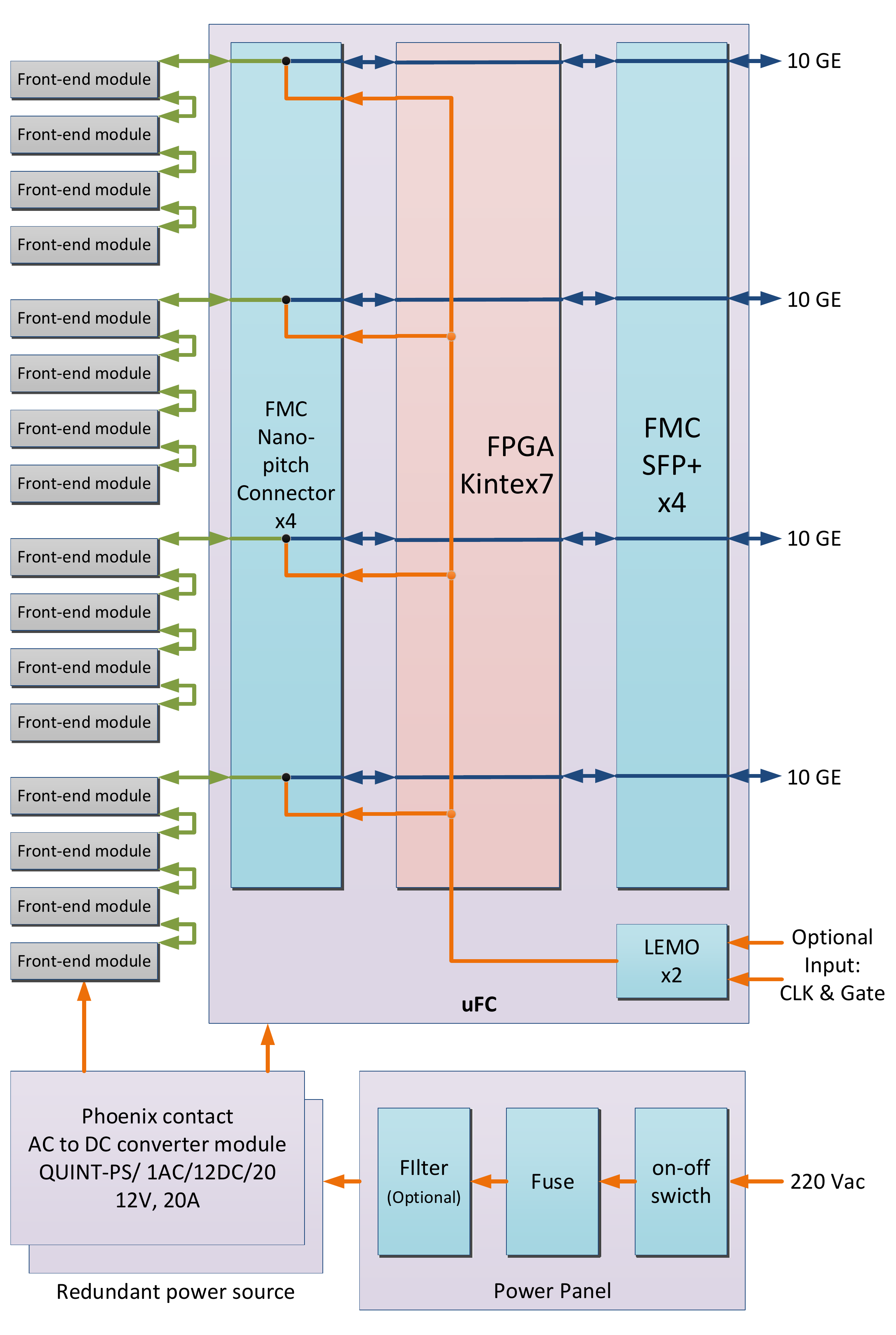}
\caption{Block diagram of HEPS-BPIX detector. The assembled front-end modules ¡°plug and play¡± into the uFC in daisy chain via Molex Nano-Pitch I/O\textsuperscript{TM} Interconnect Cable. The uFC connects to DAQ with four 10G Ethernet cables}
\label{fig_bpix_block}
\end{figure}

Figure \ref{fig_bpix_block} shows an overview of the system architecture including the 16 front-end modules and one uFC. The uFC distributed clock and trigger signals to front-end modules and forward packets between front-end module and DAQ via 10G Ethernet.

\begin{figure}[!htbp]
\centering
\includegraphics[width=2.5in]{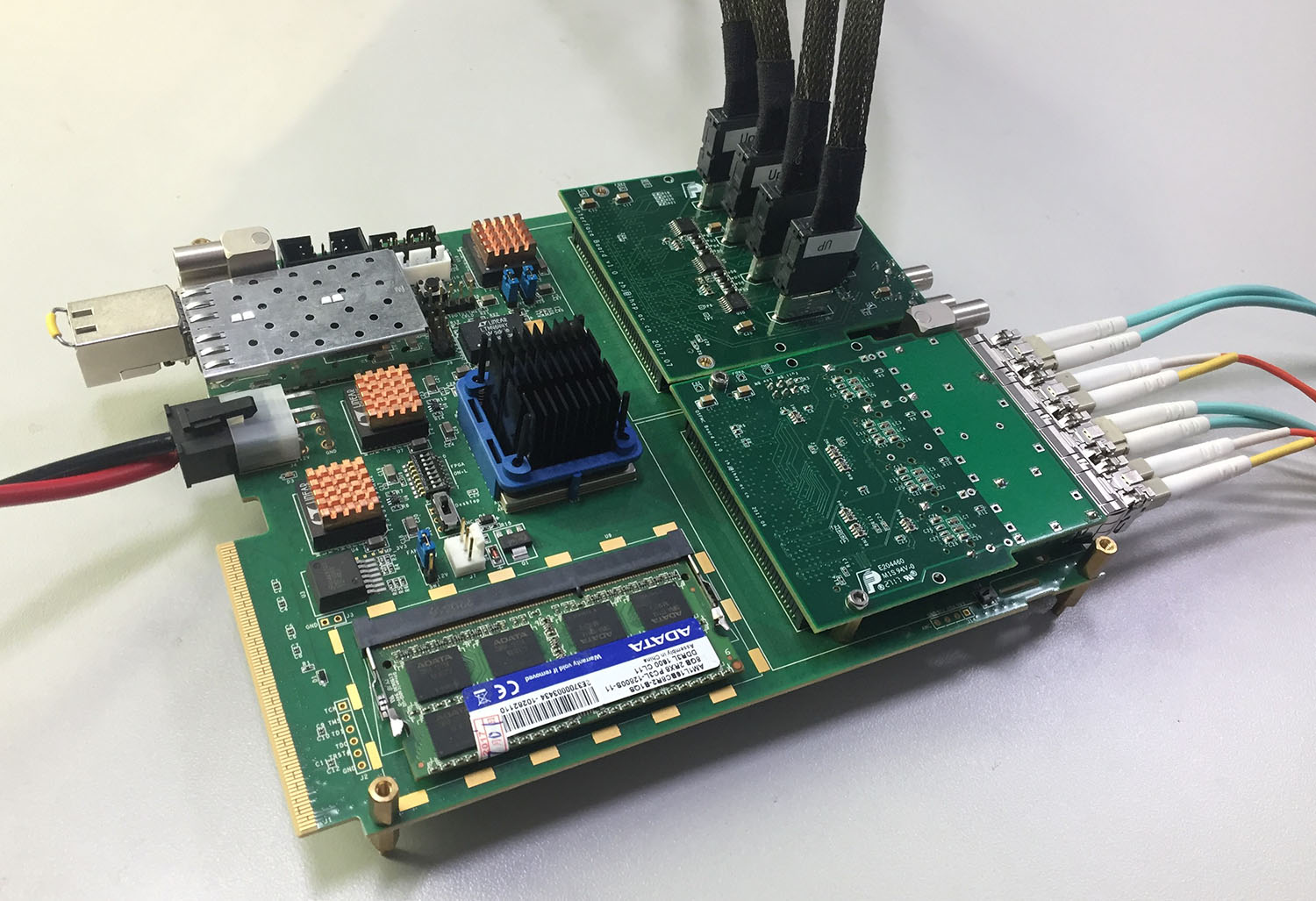}
\caption{uFC with Quad SFP/SFP+ transceiver FMC and Quad Molex Nano-Pitch I/O\textsuperscript{TM} Interconnect FMC mounted.}
\label{fig_fmc}
\end{figure}

\begin{figure}[!htbp]
\centering
\includegraphics[width=2.5in]{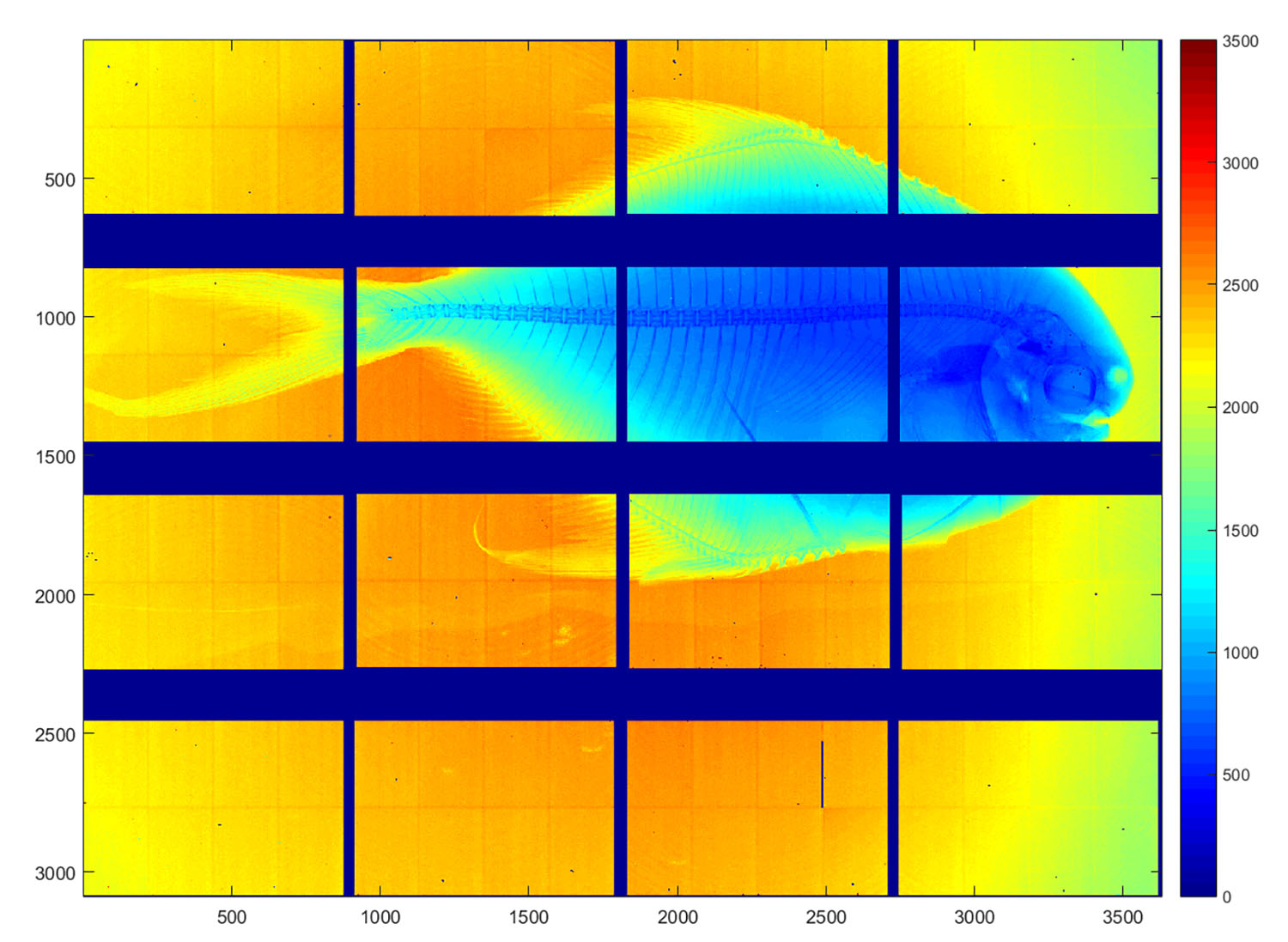}
\caption{X-Ray image of fish. The X-rays were produced by an Au tube powered at 20 kV.}
\label{fig_xray}
\end{figure}

The connectors for two FMC cards and the cooling block for the FPGA can be seen in Figure \ref{fig_fmc}. The two different FMC cards for front-end and back-end were designed. The front-end FMC had four Molex Nano-Pitch I/O\textsuperscript{TM} connectors, and each defined 9 differential pairs for 10 GE, JTAG, clock and trigger signals. The back-end FMC was equipped with four SFP+ optical transceivers connected to data server. The X-ray image taken using the second generation HEPS-BPIX system is shown in Figure \ref{fig_xray}.

A new 16M pixels detector will be released at the beginning of 2019. Built with 100 front-end modules, this detector will equip with MicroTCA crate to handle high volumes of data, even 115Gbps or more. The current 10 GE links between back-end board and DAQ will be replaced with PCI-Express interface. Thanks to the feasibility of the uFC, no hardware need to be redesign, and most firmware and software can be reused.

\section{Summary}
The uFC is an FPGA-based MicroTCA compatible AMC targeting generic system control and data acquisition in HEP experiments. The presence of the HPC FMC sockets is a big advantage as they provide additional clock signals, user-specific I/O and high-speed transceivers that can be used to extend the connectivity as well as the I/O bandwidth.
With the prototype system, we have successfully demonstrated the feasibility of the uFC, including some of its important features such as 10G transceivers, clock and trigger function.


%



\section*{Acknowledgment}
This work was supported by a grant from the National Key Research and Development Program of China (No. 2016YFA0401301).

\ifCLASSOPTIONcaptionsoff
  \newpage
\fi

\end{document}